\documentclass[12pt]{article}
\usepackage[dvips]{graphicx}
\usepackage{subfigure}

\usepackage{epsfig,amsmath,amssymb,verbatim,mathrsfs}

\def\beq{\begin{eqnarray}}
\def\eeq{\end{eqnarray}}
\def\bea{\begin{eqnarray}}
\def\eea{\end{eqnarray}}

\newcommand{\uv}{\text{\tiny UV}}

\newcommand{\s}{\text{\tiny S}}
\newcommand{\h}{\text{\tiny H}}

\newcommand{\sm}{\text{\tiny SM}}

\newcommand{\be}{\begin{equation}}
\newcommand{\ee}{\end{equation}}

\begin{document}

\setlength{\baselineskip}{0.2in}


\begin{titlepage}
\noindent
\flushright{October 2013}
\vspace{0.2cm}

\begin{center}
  \begin{Large}
    \begin{bf}
Poincar\'e Protection for a Natural Electroweak Scale
     \end{bf}
  \end{Large}
\end{center}

\begin{center}

\begin{large}

{Robert~Foot,$^{\dagger}$\footnote{rfoot@unimelb.edu.au} Archil~Kobakhidze,$^{*}$\footnote{archilk@physics.usyd.edu.au} Kristian~L.~McDonald$^{*}$\footnote{klmcd@physics.usyd.edu.au} and Raymond~R.~Volkas$^{\dagger}$}\footnote{raymondv@unimelb.edu.au}\\
     \end{large}
\vspace{0.5cm}
  \begin{it}
ARC Centre of Excellence for Particle Physics at the Terascale\\
\vspace{0.5cm}
($\dagger$) School of Physics, The University of Melbourne, VIC 3010, Australia\\
\vspace{0.5cm}
($*$) School of Physics, The University of Sydney, NSW 2006, Australia\\\vspace{0.5cm}

\end{it}

\end{center}


\begin{abstract}
We discuss a class of technically-natural UV extensions of the Standard Model in which the electroweak scale is shielded from large radiative corrections from heavy UV physics due to an enhanced Poincar\'e symmetry. Such  heavy sectors can be invoked to provide solutions to known shortcomings of the Standard Model, such the strong-CP problem, the absence of dark matter, and the lack of neutrino masses. We  discuss the relationship to scale-invariant models.

\end{abstract}

\vspace{1cm}

\end{titlepage}

\setcounter{footnote}{0}
\setcounter{page}{1}


\vfill\eject


\section{Introduction}
The concept of naturalness, as it pertains to the hierarchy problem~\cite{Gildener:1976ai,Susskind:1978ms}, has played  a key role in particle physics research in recent decades. However, the frameworks most often invoked to resolve the apparent naturalness issues have  thus far failed to manifest at the Large Hadron Collider (LHC). It is therefore timely to reexamine our concept of naturalness and consider what lessons may be extracted from a null result at the LHC.\footnote{Null result here refers to the non-observation of particles associated with conventional methods of protecting the weak scale, like supersymmetric particles. It may include other TeV scale  (or below) discoveries not connected with the hierarchy problem.} 

 There appear to be three main ways to react to a null result at the LHC. The first is to hypothesize that our concerns regarding  naturalness are largely correct but that Nature is \emph{not completely} natural. From this perspective the arguments in favor of new TeV scale physics largely hold, but not completely, and the new physics associated with  mainstream solutions to the hierarchy problem should manifest at some energy scale beyond the TeV scale. For example, if the supersymmetry (SUSY) partners have masses of order 10~TeV, large corrections to the Higgs mass from deep ultraviolet (UV) scales would largely, but not completely, cancel out. Such theories would be protected from severe naturalness issues between the weak scale and deep UV scales but would retain some amount of tuning --- they would be ``almost, but not quite, natural."

A second, more extreme possibility is that our concept of naturalness was erroneously applied to the electroweak scale. For example, if the Higgs mass is an environmental quantity that assumes different values in distinct regions of a multiverse, the small observed value could be anthropically selected; that is, the Higgs mass is small simply because we would not be here otherwise. This scenario, despite being scientifically unfortunate, remains possible.

The third possibility is that our arguments regarding naturalness are in fact correct and that  the UV completion of the Standard Model (SM) possesses a mechanism that protects the weak scale. This appears to be the least extreme case, in the sense that it maintains consistency with the dominant philosophies prior to the LHC era. Certainly it is worth considering this possibility and asking what insights the LHC may provide. In this work we discuss a class of UV extensions of the SM that are consistent with this view and in which the electroweak scale is protected by an enhanced Poincar\'e symmetry. This class of models admits technically-natural hierarchical scales that can be invoked to address known shortcomings of the SM such as the strong-CP problem, the absence of dark matter, and the lack of neutrino masses.

The paper is structured as follows. In Sec.~\ref{sec:light_scalar} we discuss the basic hierarchy problem for a light scalar field and consider  example UV completions in which the light scalar can be protected by an enhanced Poincar\'e symmetry. The application of these ideas to the SM is discussed in Sec.~\ref{sec:SM}, with some specific examples given in Sec.~\ref{sec:example_axion}. In Sec.~\ref{sec:poincare_or_SI} we discuss the role that Poincar\'e protection can play in scale-invariant theories, and we conclude in Sec.~\ref{sec:conc}.
\section{Technically Natural UV Completions for a Light Scalar\label{sec:light_scalar}}
To motivate the class of UV completions of present interest, we first discuss the quadratic sensitivity of a scalar mass to a hard UV cutoff. Consider a low-energy theory describing a self-interacting scalar field $S$, with the Lagrangian
\bea
\mathcal{L}_S&=& \partial^\mu S^*\partial_\mu S-m_0^2|S|^2 - \lambda_S |S|^4 \ .
\eea
The hierarchy problem is manifest by, for example, the quadratic sensitivity of the mass corrections when  the divergent loop integrals are regularized  with a hard cutoff $\Lambda$,\footnote{To reduce clutter we suppress numerical loop-factors throughout.}
\bea
m_S^2&=&m_0^2+\delta m_S^2\ \equiv\ m_0^2 +  \lambda_S \{\Lambda^2 + m_0^2 \log(m_0^2/\Lambda^2)\}\ ,\label{light_mass_Ir_theory}
\eea
where $m_S$ is the physical scalar mass. If there are no physical scales in the UV  the $\Lambda$ dependence is  viewed merely as an artifact of regularization that is  devoid of physical content; one can simply renormalize the theory to trivially remove the $\Lambda$-dependence.\footnote{In this case the softly-broken classical scale-invariance of the theory can be invoked to argue that the hard cutoff artificially breaks the symmetry structure of the theory and the quadratic divergences are unphysical~\cite{Bardeen:1995kv};~i.e.~absent two hierarchical physical scales there can be no hierarchy problem.}  However, if there are physical scales in the UV that induce the $\Lambda$-dependence of the mass correction, this effect is physical and cannot be ignored. The theory then has a  hierarchy problem for $\lambda_S\Lambda^2 \gg m_S^2$. 

Let us discuss two UV completions of this theory in which new physics  is associated with the UV scale  $\Lambda\gg m_S$. First we UV-extend the theory by adding a heavy scalar $H$, giving
\bea
\mathcal{L}_{SH}&=& \mathcal{L}_S+ \partial^\mu H^*\partial_\mu H-M_H^2|H|^2 - \lambda_{SH} |H|^2|S|^2-\lambda_H |H|^4\ .
\eea
The loop-corrected expression for the light scalar mass is now
 \bea
m_S^2&=&m_0^2 +  \lambda_S \{\Lambda_{\uv}^2 +  m_0^2 \log(m_0^2/\Lambda_{\uv}^2)\}  + \lambda_{SH}\{\Lambda_{\uv}^2+  M_H^2 \log(M_H^2/\Lambda_{\uv}^2)\}\ ,
\eea
where we cut the loop integrals off at $\Lambda_{\uv}\gg M_H$. Consider the case where this theory has no additional physical scales in the UV; that is, $\Lambda_{\uv}$ is merely a regularization tool and the quadratic sensitivity is an artifact to be subtracted during  renormalization.\footnote{The arguments of Ref.~\cite{Bardeen:1995kv} generalize to any theory with explicit masses that softly break scale invariance; provided the  UV scale is not physical, the sensitivity to a hard UV cutoff  is artificial and can be subtracted, ensuring that classical scale-invariance is restored if the masses are sent to zero.  The seemingly \emph{ad hoc} neglect of quadratic divergences in the ``finite naturalness'' approach of Ref.~\cite{Farina:2013mla} can be justified by softly-broken scale invariance.} However, even if the apparent  hierarchy problem with respect to $\Lambda_{\uv}$ is unphysical, the sensitivity of $m_S$ to the heavy mass scale $M_H$ persists and a genuine hierarchy problem remains.\footnote{This was how the hierarchy problem was first identified: via the calculable  radiative dependence of a light mass-scale on  a heavy particle~\cite{Gildener:1976ai}.}

This $M_H$-sensitivity is the origin of the hierarchy problem observed in Eq.~\eqref{light_mass_Ir_theory}. It provides the physical justification for attributing a hierarchy problem to Eq.~\eqref{light_mass_Ir_theory} despite the presence of bare parameters; that is, the $\Lambda^2$ term in Eq.~\eqref{light_mass_Ir_theory} encodes the effects of the heavy field $H$. The UV completion teaches us that $\Lambda^2 \sim (\lambda_{SH}/\lambda_S)\times M^2_H$ and allows one to precisely define the hierarchy problem: for a (renormalized) coupling of $\lambda_S\sim\mathcal{O}(1)$ a hierarchy problem arises if $\lambda_{SH}M^2_H\gg m_S^2$.

Now consider a second UV completion of the theory described by $\mathcal{L}_S$. Let the UV completion contain a heavy fermion that Yukawa couples to $S$, rather than the heavy scalar $H$. The Lagrangian is
\bea
\mathcal{L}_{S\psi}&=& \mathcal{L}_S+  \left( i\overline{\psi}\gamma^\mu\partial_\mu \psi-M_\psi\overline{\psi}\psi - \lambda_{\s\psi} \overline{\psi}\psi S +\mathrm{H.c.}\right)\ ,
\eea
and the loop-corrected expression for the light scalar mass is now
 \bea
m_S^2&=&m_0^2 +  \lambda_S \{\Lambda_{\uv}^2 + m_0^2 \log(m_0^2/\Lambda_{\uv}^2)\}  - \lambda_{\s\psi}^2 \{\Lambda_{\uv}^2 + M_\psi^2 \log(M_\psi^2/\Lambda_{\uv}^2)\}\ , \label{fermion_loop_mass}
\eea
where $\Lambda_{\uv}\gg M_\psi$. We again consider the case with no additional fields in the UV so the $\Lambda_{\uv}$-dependence is merely an artifact that can be removed. As with the heavy-scalar example, a hierarchy problem exists due to the dependence of Eq.~\eqref{fermion_loop_mass} on the  heavy fermion mass $M_\psi$, requiring a tuning between $m_0$ and $M_\psi$ to maintain $m_S\ll M_\psi$. In this case the $\Lambda$-dependence of Eq.~\eqref{light_mass_Ir_theory} encodes the existence of $\psi$ in the UV, and we have $\Lambda^2\sim(\lambda_{\s\psi}^2/\lambda_{S})\times M_\psi^2$. A hierarchy problem emerges if  $\lambda_{\s\psi}^2M^2_\psi\gg m_S^2$ for $\lambda_S\sim\mathcal{O}(1)$, as the radiative corrections become much larger than the physical mass $m_S$.

Both of these UV completions reveal the standard hierarchy problem --- light scalars that couple to heavy UV fields are unnatural. The supersymmetric solution to this problem invokes a new symmetry to ensure the $\Lambda$-dependence of Eq.~\eqref{light_mass_Ir_theory}  cancels out amongst particles and their SUSY partners. However, the above examples reveal another symmetry that can protect the light scalar mass in a class of UV completions of $\mathcal{L}_S$. 

The troublesome $\Lambda^2$ term encodes a sensitivity to both the heavy mass $M_{\psi/H}$ and a dependence on the mixing coupling $\lambda_{\s\psi/\h}$. The UV completions show that one can alleviate, and eventually turn off, the hierarchy problem by taking the limit $\lambda_{\s\psi/\h}\rightarrow0$. The  total Lagrangian can be written as
\bea
\mathcal{L}_{\s\psi/\h}&=& \mathcal{L}_S+ \mathcal{L}_{\rm mix} +\mathcal{L}_{\psi/\h}\ ,
\eea
where the mixing Lagrangian contains the $\lambda_{\s\psi/\h}$ term.  In the limit $\lambda_{\s\psi/\h}\rightarrow0$ this Lagrangian decouples to give two independent sectors:
\bea
\mathcal{L}_{\s\psi/\h}&\longrightarrow& \mathcal{L}_S +\mathcal{L}_{\psi/\h}\quad\mathrm{for}\quad\lambda_{\s\psi/\h}\rightarrow 0\ .\label{eq:decoupled_lagrangians}
\eea
At the level of the action,
\bea
S = S_S + S_{\psi/\h} = \int d^4x\, \mathcal{L}_S(x) + \int d^4x'\, \mathcal{L}_{\psi/\h}(x')\ ,
\eea
where the coordinate dependences of the Lagrangians are understood to be acquired through the fields and their first derivatives.  Note that the action separates into integrals with independent integration variables.  It is thus clear that in this limit one can  perform independent Poincar\'e transformations that leave $S_S$ and $S_{\psi/H}$ separately invariant, so that the symmetry of the theory is enhanced to the product group $\mathcal{G}_P^{\s}\otimes \mathcal{G}_P^{\psi/\h}$ ($ \mathcal{G}_P$ denotes the Poincar\'e group). Turning on small nonzero values of $\lambda_{\s\psi/\h}$ breaks the product group to the diagonal subgroup, 
\bea
\mathcal{G}_P^\s\otimes \mathcal{G}_P^{\psi/H}&\longrightarrow&\mathcal{G}_P^{\s+\psi/\h}\ .
\eea
Small values of $\lambda_{\s\psi/\h}$ are thus technically natural due to an enhanced Poincar\'e symmetry  in the limit $\lambda_{\s\psi/\h}\rightarrow0$; this is reflected in the beta-functions for $\lambda_{\s\psi/\h}$, which  have a fixed point at $\lambda_{\s\psi/\h}=0$. However, small values of $\lambda_{\s\psi/\h}\rightarrow0$ also turn off the radiative corrections to the light scalar mass from the heavy UV physics. Thus, the symmetry-enhancing limit corresponds to the limit in which  a naturally light scalar emerges --- an enhanced Poincar\'e symmetry can protect a light scalar. Light scalars are  therefore technically natural if the UV completion contains weakly-coupled heavy physics with mass of $\mathcal{O}(M)$, provided  the UV physics decouples from the light-sector in the limit $\lambda_{\rm mix}\rightarrow0$, and the mixing coupling (or couplings) satisfies $\lambda_{\rm mix}\lesssim\mathcal{O}(m_S/M)$.\footnote{ The discussion of this section can also be framed in terms of the beta functions for the parameters of theory, as we outline in the Appendix.}

Note that the enhanced Poincar\'e symmetry  reflects  an increase  the number of independently conserved stress-energy tensors in the theory. With $\lambda_{\s\psi/\h}\ne0$ there is a single conserved stress-energy tensor  that describes both the light and heavy sectors: $\partial_\mu T_{\s+\psi/\h}^{\mu\nu}=0$. This conserved current reflects the Poincar\'e symmetry $\mathcal{G}_P^{\s+\psi/\h}$ of the theory. In the limit $\lambda_{\s\psi/\h}\rightarrow0$, the two sectors decouple and the corresponding stress-energy tensors are independently conserved: $\partial_\mu T^{\mu\nu}_{\s}=\partial_\mu T^{\mu\nu}_{\psi/\h}=0$. Accordingly one can identify two exact symmetries, namely $\mathcal{G}_P^\s\otimes \mathcal{G}_P^{\psi/H}$. Turning on $\lambda_{\s\psi/\h}\ne0$ breaks this symmetry back to the diagonal group and gives a single conserved stress-energy tensor.

 Poincar\'e symmetry is only an approximate local symmetry in general relativity.  In the presence of gravity, independent Poincar\'{e} symmetries are not obtained in the $\lambda_{\rm mix} \to 0$ limit, even as approximate local symmetries, because gravity mixes the two sectors.  The question of whether or not gravitational interactions destabilize the hierarchy cannot be answered without an explicit theory of quantum gravity that is combined with the non-gravitational physics.  It could be that Newton's constant acts merely as a small coupling, so the gravitational breaking of the independent Poincar\'{e} symmetries leads only to small corrections that do not destroy the hierarchy \cite{Strumia}.  This is not guaranteed, but it is conceivable.  In any case, it is clear that Poincar\'{e} protection is rigorously correct for the non-gravitational sectors.

\section{Poincar\'e Protected UV Extensions of the Standard Model\label{sec:SM}}

The above ideas can be applied to the SM to motivate a class of technically-natural UV extensions. Denote the Lagrangian for the SM as $\mathcal{L}_{\sm}$.\footnote{This Lagrangian could also describe yet-to-be-discovered new physics, with masses at the TeV scale or less, that does not destabilize the weak scale (i.e.~radiative corrections from fields in $\mathcal{L}_{\sm}$ do not create a hierarchy problem).} Let the field-theoretic UV completion of the low-energy  theory described by $\mathcal{L}_{\sm}$ contain $n$ additional sectors, described by the Lagrangians $\mathcal{L}_i$, $i\in\{1,\,2\,...,\,n\}$. These sectors mix with the SM via the Lagrangian $\mathcal{L}^{i}_{\rm mix}$.\footnote{In general there could  be Lagrangians $\mathcal{L}_{ij}$ that mix the additional sectors. However, for  low-energy purposes one can redefine the sectors, without loss of generality, such that multiple sectors and their mixing are described by a given Lagrangian $\mathcal{L}_i$.} The complete Lagrangian for the UV theory is then
\bea
\mathcal{L}&=& \mathcal{L}_{\sm} + \sum_i \left( \mathcal{L}^{i}_{\rm mix}+\mathcal{L}^{i} \right) \ .
\eea
Consider one of these sectors, $\mathcal{L}^j$. The Lagrangian $\mathcal{L}_{\rm mix}^j$ contains  a set of mixing couplings $\lambda_{\rm mix}^{j}$, such that the limit $\lambda_{\rm mix}^j\rightarrow 0$ decouples the sector $\mathcal{L}^j$ and increases the symmetry of the theory:
\bea
\mathcal{G}_P^{\sm+n}&\longrightarrow&\mathcal{G}_P^{\sm+(n-1)}\otimes \mathcal{G}_P^{j}\quad\mathrm{for}\quad\lambda_{\rm mix}^j\rightarrow 0\ .
\eea
The small $\lambda^j_{\rm mix}$ limit is therefore technically natural, and  provided $\lambda^j_{\rm mix}\lesssim \mathcal{O}(m_h/M_j)$ the weak scale is protected from destabilizing radiative corrections from the heavy fields with $\mathcal{O}(M_j)$ masses  in the sector $\mathcal{L}^{j}$.

This procedure can be repeated to show that the weak scale is protected from radiative corrections involving the heavy fields in a given sector $\mathcal{L}^i$, provided $\lambda^i_{mix}\lesssim \mathcal{O}(m_h/M_i)$, due to the enhanced symmetry
\bea
\mathcal{G}_P^{\sm+n}&\longrightarrow&\mathcal{G}_P^{\sm}\otimes \mathcal{G}_P^{1}\otimes \mathcal{G}_P^{2}\otimes \ldots\otimes \mathcal{G}_P^{n}\quad\mathrm{for}\quad\lambda_{\rm mix}^i\rightarrow 0\quad\forall \ i\ .
\eea
In this way, one arrives at a technically natural UV extension of the SM, comprised of $n$ weakly-coupled heavy sectors in addition to the light SM sector. The  heavy sectors can be invoked to provide solutions to known shortcomings of the SM. 
\section{An Example: The (Technically Natural) Invisible Axion Model\label{sec:example_axion}}
 The invisible axion model provides a simple example of a UV extension of the SM in which  radiatively-stable hierarchical scales can be understood in terms of an enhanced Poincar\'e symmetry~\cite{Volkas:1988cm}.  This framework addresses the strong CP problem via the Peccei-Quinn (PQ) mechanism~\cite{Peccei:1977hh} while largely hiding the axion~\cite{Weinberg:1977ma} from experimental searches. The details of the model are well known to the literature; one extends the SM to include a second scalar doublet $H_2\sim(1,2,1)$ and  a gauge-singlet scalar $N\sim(1,1,0)$, and imposes a global $U(1)$ Peccei-Quinn symmetry~\cite{Kim:1979if}. The axion is ``invisible" if it resides primarily in the singlet-scalar $N$, and successful phenomenology follows when the PQ symmetry is broken by a nonzero vacuum expectation value (VEV) for $N$, provided the VEV is in the range  $10^8~\mathrm{GeV}\lesssim \langle N\rangle \lesssim10^{12}~\mathrm{GeV}$. In addition to the axion, the spectrum contains a heavy scalar with mass $M_N=\mathcal{O}(\langle N\rangle)$, which greatly exceeds the weak scale. 

Successful electroweak symmetry breaking requires $\langle H_{1}\rangle^2 +\langle H_2\rangle^2 \simeq (174~\mathrm{GeV})^2$, mandating the hierarchy $\langle H_{1,2}\rangle/\langle N\rangle\ll1$. Our interest is in the  radiative stability of this hierarchy. The Lagrangian can be written as
\bea
\mathcal{L}&=&\mathcal{L}_{SM+H_2} +\mathcal{L}_{\rm mix} +\mathcal{L}_N\ ,
\eea
where the first term describes the SM sector (plus $H_2$) and $\mathcal{L}_N$ contains the heavy sector. Communication between the two sectors is controlled by the mixing Lagrangian:
\bea
-\mathcal{L}_{mix} &=& \lambda_{1N} |H_1|^2 |N|^2 +\lambda_{2N}|H_2|^2|N|^2 +\kappa H_1^\dagger H_2N^2 +\mathrm{H.c.}
\eea
The hierarchy $\langle H_{1,2}\rangle/\langle N\rangle\ll1$ can be preserved at tree-level if the mixing couplings are suppressed:
\bea
\lambda_{1N},\,\lambda_{2N},\, \kappa &\lesssim& \frac{(10^2~\mathrm{GeV})^2}{M_N^2}\ \ll\ 1\ .
\eea
Small values of $\lambda_{1N},\,\lambda_{2N},\,\kappa\ll1$ are technically natural because the two sectors decouple in the limit $\lambda_{1N},\,\lambda_{2N},\,\kappa\rightarrow0$, giving the enhanced Poincar\'e symmetry $\mathcal{G}_P^{SM+H_2}\otimes \mathcal{G}_P^N$~\cite{Volkas:1988cm}. Radiative corrections to light scalar masses from the heavy sector are also controlled by these couplings. In the above parameter space, radiative corrections do not destabilize the hierarchy:
\bea
\left\{
\begin{array}{c}
\delta m_1^2 \\
\delta m_{12}^2\\
\delta m_2^2 
\end{array}
\right\}
&\sim&
\left\{
\begin{array}{c}
\lambda_{1N}  \\
  \kappa  \\
 \lambda_{2N} 
\end{array}
\right\}\times M_N^2
\ \lesssim\ \mathcal{O}(100~\mathrm{GeV})^2\ .
\eea
This model solves the strong CP problem of the SM  and exemplifies the use of a technically-natural hierarchy of physical scales that is protected by an enhanced Poincar\'e symmetry. Furthermore, the axion can also play the role of dark matter~\cite{Preskill:1982cy}. 

Another simple example demonstrating our ideas is the standard seesaw mechanism with Majorana masses for the right-handed neutrinos restricted to the range $M_R\lesssim 7\times10^7$~GeV. For these values, the Dirac Yukawa couplings between SM neutrinos and the right-handed neutrinos must be small, $y_\nu\ll1$, and radiative corrections to the weak scale from loops containing neutrinos do not destabilize the weak scale~\cite{Vissani:1997ys,Farina:2013mla}. In the limit that these Yukawa couplings  vanish, $y_\nu\rightarrow0$, the Poincar\'e symmetry of the model is enhanced to $\mathcal{G}_P^{\sm}\otimes \mathcal{G}_P^{\nu_R}$, providing a technically-natural interpretation for the radiative stability of the hierarchically separated scales. Other UV extensions of the SM are of course possible.

\section{Poincar\'e Protection or Scale Invariance?\label{sec:poincare_or_SI}}

Scale invariance has recently gained  attention as a symmetry that may play a role in protecting the electroweak scale~\cite{Meissner:2006zh,Foot:2007as,Foot:2007iy}. Classically  scale-invariant extensions of the SM  have been considered~\cite{Iso:2009ss,Foot:2010av,AlexanderNunneley:2010nw}, and a class of completely scale-invariant theories (i.e.~the scale invariance holds even at the quantum level) have also been invoked~\cite{Shaposhnikov:2008xi}. Here we discuss the connection between scale-invariant models and Poincar\'e protection~\cite{Foot:2007iy,Foot:2010av}. We also clarify  recent statements regarding the inability of scale-invariant models to protect the weak scale~\cite{Tamarit:2013vda}.

Our main points can be illustrated with a simple classically scale-invariant toy model, comprised of the gauge symmetry $SU(2)_1\otimes SU(2)_2$, and two scalar doublets, $\Phi_1\sim(2,1)$ and $\Phi_2\sim(1,2)$. The most general scale-invariant  potential  is
\bea
V(\Phi_1,\Phi_2)&=&\frac{\lambda _1}{2}(\Phi_1^\dagger\Phi_1)^2 -\lambda _{\rm mix}(\Phi_1^\dagger\Phi_1) (\Phi_2^\dagger\Phi_2) +\frac{\lambda _2}{2}(\Phi_2^\dagger\Phi_2)^2\ . 
\eea
The quartic couplings are running parameters that depend on the renormalization scale $\mu$. We consider the parameter space with $\lambda_{\rm mix}\ll1$ and, without loss of generality,  consider the case where $\lambda_2(\mu=\mu_2)=0$ at a UV scale   $\mu_2\gg \mu_1$, where $\lambda_1(\mu=\mu_1)=0$. In this case radiative corrections trigger a nonzero VEV, $\langle\Phi_2\rangle\ne0$, breaking the $SU(2)_2$ symmetry via the Coleman-Weinberg mechanism~\cite{Coleman:1973jx} at the UV scale $\mu_2$. This symmetry breaking also gives a negative mass-squared for $\Phi_1$, which induces the VEV
\bea
\langle\Phi_1\rangle^2&=& \frac{\lambda_{\rm mix}}{\lambda_1}\langle\Phi_2\rangle^2\ ,\label{eq:light_vev}
\eea
and breaks the $SU(2)_1$ symmetry. 

For $\lambda_{\rm mix}\ll1$ the  model contains two hierarchically separated physical scales, $\langle\Phi_1\rangle\ll\langle\Phi_2\rangle$.  The heavy sector consists of three massive vectors with mass  $M_2=\mathcal{O}(\langle \Phi_2\rangle)$, and a physical scalar $\phi_2$ whose mass is  suppressed relative to $M_2$ by a loop factor (this is the dilaton or scalon~\cite{Coleman:1973jx,Gildener:1976ih}). The  light sector is comprised of a physical scalar $\phi_1$ and three vectors, all with $\mathcal{O}(\langle\Phi_1\rangle)$ masses. Denoting the light scalar mass by $m_1$, radiative corrections to this mass from the heavy sector are of order $\delta m_1^2\sim \lambda_{\rm mix} M_2^2=\mathcal{O}(m_1^2)$, which are automatically small.  Thus, the light scalar is protected from large radiative corrections and the hierarchical scales are radiatively stable.

In addition to the scale-invariance, this toy model possesses an enhanced Poincar\'e symmetry in the limit $\lambda_{\rm mix}\rightarrow0$. One would like to know if it is the scale invariance  or the increased Poincar\'e symmetry  that ensures the radiative stability of the hierarchical scales. It is easy to show that it is the latter. Let as add an additional singlet scalar $\Phi'\sim(1,1)$ to the model, giving the potential
\bea
V(\Phi_1,\Phi_2,\Phi')&=&V(\Phi_1,\Phi_2)-\lambda _{\rm mix}'(\Phi_1^\dagger\Phi_1) \Phi'^2+\lambda _{m,2}(\Phi_2^\dagger\Phi_2) \Phi'^2+\frac{\lambda _2'}{2}\Phi'^4\ .
\eea
Take the new couplings to be $\mathcal{O}(1)$ and consider the parameter space where $\Phi_2$ again develops a large VEV via dimensional transmutation while $\langle\Phi'\rangle=0$.  The nonzero VEV $\langle\Phi_2\rangle$ again triggers a VEV for $\Phi_1$ via Eq.~\eqref{eq:light_vev}, with $\langle\Phi_1\rangle\ll\langle\Phi_2\rangle$ for $\lambda_{\rm mix}\ll1$. The heavy sector now contains an additional heavy scalar ($\Phi'$) which acquires a large mass of $\mathcal{O}(M_2)$ for $\lambda_{m,2}\sim\mathcal{O}(1)$. This scalar couples to the light scalar $\phi_1$ through the $\lambda_{\rm mix}'$ term, inducing radiative corrections to the light scalar mass that are controlled by
\bea
\delta m_1'^2\sim\lambda_{\rm mix}'M_2^2=\mathcal{O}(M_2^2)\gg m_1^2\ .
\eea
Thus the radiative stability of the light scale is destroyed by the presence of the heavy scalar $\Phi'$ with $\mathcal{O}(1)$ couplings to both the heavy and light sectors. This extended toy model possesses the same (classical) scale-invariance of the original toy model, demonstrating that the scale invariance alone does not guarantee radiative stability of the light scale. Note that the extended Poincar\'e symmetry $\mathcal{G}_P^1\otimes\mathcal{G}_P^2$ is now broken by  the large coupling $\lambda_{\rm mix}'\gg\lambda_{\rm mix}$. The limit $\lambda_{\rm mix}\rightarrow0$ no longer decouples the two sectors so the light scale is not protected by an enhanced symmetry in this limit. 

Now consider the case where $\lambda_{\rm mix}'\sim\lambda_{\rm mix}\ll1$. The extended Poincar\'e symmetry emerges for $\lambda_{\rm mix}',\lambda_{\rm mix}\rightarrow0$, so this is a technically natural region of parameter space, according to 't~Hooft's definition~\cite{'tHooft:1979bh}. Radiative corrections to the light mass from $\phi_2$ remain on the order of $\delta m_1^2\sim\lambda_{\rm mix} M_2^2=\mathcal{O}(m_1^2)$, and do not destabilize the light scale. With $\lambda_{\rm mix}'\sim\lambda_{\rm mix}$ the corrections from $\Phi'$ are now of order $\delta m_1'^2\sim\lambda_{\rm mix}'M_2^2=\mathcal{O}(m_1^2)$, which are also compatible with a stable light sector. Thus, the light sector is now protected from large corrections by the extended Poincar\'e symmetry, giving  a technically-natural theory with hierarchically-separated physical scales.

Beyond demonstrating that  classical scale invariance alone is insufficient to ensure radiative stability of the light scalar~\cite{Foot:2007iy,Foot:2010av}, these toy models demonstrate that the maximal hierarchy of radiatively-stable scales  achievable in classically scale-invariant theories is determined by the \emph{largest} parameter that mixes the light and heavy sectors  (i.e.~the parameter that most strongly breaks the extended Poincar\'e symmetry). Denoting the strongest source of enhanced Poincar\'e symmetry breaking by $\lambda_{\rm max}$, the maximal radiatively-stable hierarchy is\footnote{The power of the dimensionless ratio depends on whether the coupling is, for example, a quartic coupling or a Yukawa coupling; we quote results for a quartic scalar coupling.}
\bea
\frac{M_{\rm heavy}^2}{m^2_{\rm light}}&=&\mathcal{O}(\lambda_{\rm max}^{-1})\quad\mathrm{where}\quad\lambda_{\rm max}=\mathrm{max}\{\lambda_{\rm mix}\}\ .
\eea
In the first version of the extended toy-model the enhanced Poincar\'e symmetry is most severely broken by the coupling $\lambda_{\rm mix}'$, and the largest stable hierarchy that can be achieved is $M^2_2/m^2_1\sim\lambda'^{-1}_{\rm mix}\ll \lambda_{\rm mix}^{-1}$. In the second version with $\lambda_{\rm mix}\sim\lambda_{\rm mix}'$ the larger hierarchy of $M^2_2/m^2_1\sim\lambda_{\rm mix}^{-1}$ is radiatively stable.

These points clarify the results of Ref.~\cite{Tamarit:2013vda}. That work considered the elevation of classical scale invariance to  complete  scale invariance (i.e.~including the quantum level). By regularizing the theory with a dilaton, rather than an explicit regularization scale, and demanding that the scalar potential has a flat direction, the trace-anomaly vanishes but  running couplings still emerge when scale invariance is spontaneously broken by the dilaton VEV. Provided the dilaton is very weakly coupled to the light scalar, the running couplings largely match those obtained in conventional theories with massive scalars~\cite{Tamarit:2013vda}. Upon adding a dilaton to a theory with a scalar, the most general scale-invariant action with a flat direction imposed on the scalar potential is [see Eq.~(3.1) in  Ref.~\cite{Tamarit:2013vda}; we adopt their notation]:
\bea
S&=&\int d^4x \left\{ \frac{1}{2} \partial_\mu H\partial^\mu H + \frac{1}{2} \partial_\mu \phi\partial^\mu \phi -\frac{\lambda_S}{4!}(H^2-\zeta^2\phi^2)^2 \right\}\ .\label{eq:exact_SI_example}
\eea
Here $\phi$ is the dilaton and one requires $\zeta\ll1$ to obtain the conventional behaviour for the running couplings~\cite{Tamarit:2013vda}. Models of this type, with exact scale invariance at the quantum level, have been considered as solutions to both the cosmological constant problem and the Higgs naturalness problem~\cite{Shaposhnikov:2008xi} (for earlier work see~\cite{Antoniadis:1984kd}). 

The demand that the potential has a flat direction is equivalent to imposing  a single constraint on the three quartic scalar couplings that would otherwise appear in the most general scale-invariant potential for $H$ and $\phi$. Thus, the potential in \eqref{eq:exact_SI_example} only contains  two independent parameters. As discussed in Ref.~\cite{Tamarit:2013vda}, absent a dynamical explanation for the requisite coupling relation, imposing the  flat direction amounts to a fine-tuning amongst dimensionless couplings that is equivalent to tuning  the cosmological constant.\footnote{Note that in classically scale-invariant theories one can always choose the renormalization scale such the the classical potential has a flat direction without any tuning~\cite{Gildener:1976ih}. In models with exact scale invariance the dynamics should force the dilaton to a value for which the 
potential is flat  -- without this explanation one must tune the parameters.} Reference~\cite{Tamarit:2013vda} then argues that it is also  unnatural for the scalar $H$ to remain light. The case is made by adding another scalar $F$ with $\mathcal{O}(1)$ couplings to $H$ and $\phi$ ($\lambda_{FH}$ and $\lambda_{F\phi}$, respectively). The new scalar gets a large mass due to the $\mathcal{O}(1)$ coupling with the dilaton and, by integrating out $F$,  Ref.~\cite{Tamarit:2013vda}  shows that the tuning required for the flat direction is destabilized and the light scalar $H$ receives large mass corrections. Our analysis reveals  $(i)$ why Ref.~\cite{Tamarit:2013vda}  arrives at these conclusions, and $(ii)$ that one can in fact simultaneously preserve a radiatively stable light-scalar and the flat direction in the presence of the field $F$. We discuss these points in turn.

In Eq.~\eqref{eq:exact_SI_example} the limit $\zeta\rightarrow0$ decouples the fields $\phi$ and $H$ and increases the Poincar\'e symmetry to $\mathcal{G}_P^H\otimes\mathcal{G}_P^\phi$. Thus, small values of $\zeta$ are technically natural. Furthermore, the light scalar mass in the theory described by Eq.~\eqref{eq:exact_SI_example}  is also protected by the enhanced Poincar\'e symmetry. Adding $F$ to the theory with $\mathcal{O}(1)$ couplings to both sectors strongly breaks the $\mathcal{G}_P^H\otimes\mathcal{G}_P^\phi$ symmetry and reduces the maximum natural hierarchy between $m_h$ and  $\langle\phi\rangle$ from $\langle \phi\rangle/m_h =\mathcal{O}(\zeta^{-1})\gg1$ to  $\langle \phi\rangle/m_h =\mathcal{O}(\lambda_{HF}^{-1})=\mathcal{O}(1)$. Consequently the  light scalar is radiatively unstable and $H$ becomes heavy.

If one instead adds $F$ to the model with a coupling $\lambda_{FH}\sim \zeta$, which does not break the  $\mathcal{G}_P^H\otimes\mathcal{G}_P^\phi$ symmetry more strongly than in the original model, the light scalar  remains protected by the enhanced Poincar\'e symmetry, and a hierarchy of $\langle \phi\rangle/m_h =\mathcal{O}(\zeta^{-1})\gg1$ remains radiatively stable. Therefore the  addition of $F$ does not necessarily destabilize the light scale; it is the hard breaking of $\mathcal{G}_P^H\otimes\mathcal{G}_P^{\phi}$ in Ref.~\cite{Tamarit:2013vda} that destabilizes the hierarchy.  One can always add $F$ to the model with couplings $\lambda_{FH}\sim \zeta$ and $\lambda_{F\phi }=\mathcal{O}(1)$, such that: $F$ is heavy, the $\mathcal{G}_P^H\otimes\mathcal{G}_P^\phi$ symmetry remains weakly broken, and the potential possesses a flat direction.\footnote{This is readily deduced from Ref.~\cite{Gildener:1976ih}. Consider the classically scale-invariant potential for $H$, $F$ and $\phi$, and choose the renormalization scale $\mu=\mu_{\rm flat}$, at which the potential has a flat direction. Now take the technically-natural limit of small $\lambda_{FH}$ and $\lambda_{H\phi}$, which preserves the flat direction. The coupling relations applicable at the scale $\mu_{\rm flat}$ can be used to  define the single fine-tuning necessary to generate a scalar potential with a flat direction in the theory with exact scale invariance, while preserving the properties mentioned in the text.} Thus, the only tuning required is the one giving a flat direction and provided the $\mathcal{G}_P^H\otimes\mathcal{G}_P^\phi$ symmetry is not strongly broken a naturally light scalar persists. It is not surprising that  adding  field-theoretic UV physics that strongly breaks the enhanced Poincar\'e symmetry  destabilizes the hierarchy --- this is similar to adding  a non-complete SUSY multiplet in the UV to re-introduce tuning in a SUSY model.

If the weak scale originates  from a scale-invariant theory, be it classical or exact,  with a single source of symmetry breaking (e.g.~a dilaton VEV or via a single occurrence of dimensional transmutation), one expects that either: ($a$) all scales are of a similar magnitude because the couplings  involved are non-hierarchical~\cite{Foot:2007as}, or ($b$) some technically-natural  small couplings exist, which in turn generate technically-natural hierarchical  scales~\cite{Foot:2007iy,Foot:2010av}. In the first case the new physics can be within reach of the LHC, while  the scale of new physics is less clear in the second. Both cases are technically natural and simply correspond to different regions of parameters space. To use the example of Ref.~\cite{Tamarit:2013vda}, one can break scale invariance and obtain $m_H\sim m_F\sim\langle \phi\rangle$ if all  couplings are $\mathcal{O}(1)$, or have technically-natural hierarchies like $m_H\sim m_F\ll\langle \phi\rangle$, or $m_H \ll m_F\sim\langle \phi\rangle$, corresponding to the enhanced Poincar\'e symmetry $\mathcal{G}_P^{H+F}\otimes\mathcal{G}_P^\phi$, or $\mathcal{G}_P^H\otimes\mathcal{G}_P^{\phi+F}$, respectively, if the parameters are  hierarchical.

In the event that Nature employs small couplings to shield the weak scale from UV physics, one would ultimately like to  understand why. However, provided they are technically natural, it is not mandatory for the low-energy theory to explain their origin. Explaining the small couplings  likely requires one to understand the origin of coupling constants in general, which is a non-trivial matter. One could  imagine, however,  that some couplings result from a type of tunneling or instanton effect in the ultimate UV theory, in which case the emergence of small couplings linking otherwise disconnected sectors would not be surprising.




\section{Conclusion\label{sec:conc}}
The LHC has thus far failed to find evidence for TeV scale physics that protects the weak scale from large UV corrections.  This empirical fact motivates one to reevaluate commonly held views on naturalness and to consider alternative UV completions of the  SM. In this work we discussed a class of models in which the weak scale is protected from large radiative corrections due to  an enhanced Poincar\'e symmetry. This approach allows heavy UV physics that can be invoked to explain known empirical shortcomings of the SM, and is consistent with both a null result at the LHC and  our long-held views on (technical) naturalness.

\section*{Acknowledgments\label{sec:ackn}}
This work was supported by the Australian Research Council.
\appendix
\section{Naturalness and RGEs}
The discussion in the text can be framed in terms of the renormalization group equations (RGEs). In the theory describing just the light scalar $S$, the scalar mass softly breaks  classical scale-invariance. This implies that a regularization scheme can be found for which the  renormalized scalar mass $m_S(\mu)$ has a fixed point at $m_S\rightarrow0$. This scheme is of course dimensional regularization (DR),  which gives
\bea
\frac{\partial m^2_S(\mu)}{\partial t}&\propto& \lambda_S(\mu)\, m_S^2(\mu)\ ,\label{eq:light_scalar_rge}
\eea
and displays the fixed point as expected (here $t=\ln\mu$ and we suppress numerical factors). In this scheme it is clear that an initially small value of the scalar mass stays small under RGE running, and is therefore natural. 
This information alone, however, is not enough to tell us with certainty that the light scalar is natural in the presence of new UV physics; 
it only tells us that the light scalar may be natural if the UV physics is such that the fixed-point behaviour of Eq.~\eqref{eq:light_scalar_rge} is not spoilt; i.e.~it depends on the nature of the UV physics.  This is clear in other regularization schemes, where the RGE contains extra terms on the right-hand side that spoil the IR fixed point in Eq.~\eqref{eq:light_scalar_rge}. All one can conclude is that the light scalar may or may not be natural, depending on the details of the UV physics; one cannot say for certain that the theory has a hierarchy problem.

Next add the heavy scalar $H$ to  the theory. Classical scale invariance is now softly broken by both scalar masses and is restored in the limit $M_H,\, m_S\rightarrow0$. Being softly-broken, there again exists a regularization scheme in which the running of the dimensionful parameters has a fixed point in the limit $M_H,\, m_S\rightarrow0$. Thus, in DR the RGE for the light scalar mass is now
\bea
\frac{\partial m^2_S(\mu)}{\partial t}&\propto& \lambda_S(\mu)\, m_S^2(\mu)\ +\ \lambda_{SH}(\mu)\, M_H^2(\mu)\ ,\label{eq:light_heavy_scalar_rge}
\eea
which displays the fixed point for $M_H,\,m_S\rightarrow0$. This informs us that the quadratic divergences may not be physical and  may simply reflect the use of a scheme that inaccurately reflects the symmetry structure of the theory; i.e.~one that breaks scale invariance in a hard way that is not representative of the quantum breaking of scale invariance expected from the Ward identities. The argument of  Ref.~\cite{Bardeen:1995kv} can be invoked  to justify the neglect of the quadratic divergences, depending on the nature of the UV physics. However, for $M_H^2\gg m_S^2$ there is a hierarchy problem that cannot be removed with a choice of regularization scheme; if the masses are hierarchical and the couplings are $\mathcal{O}(1)$, the light scalar mass  rapidly runs to large values unless it is fine-tuned. The physical content of this hierarchy problem cannot be ignored.

Equation~\eqref{eq:light_heavy_scalar_rge} shows that another fixed point emerges in the theory with two scalars; the simultaneous limit $m_S^2\rightarrow0$ and $\lambda_{SH}\rightarrow0$ also gives a fixed point. Thus, the hierarchy $m_S^2/M_H^2\ll1$ can be preserved under RGE running provided values  of $\lambda_{SH}\ll1$ are technically natural. This is precisely what happens if the limit $\lambda_{SH}\rightarrow0$ enhances the Poincar\'e symmetry of the theory: radiative corrections to $\lambda_{SH}$ are controlled by $\lambda_{SH}$ and the relevant RGE  displays a fixed point at $\lambda_{SH}\rightarrow0$:
\bea
\frac{\partial \lambda_{SH}(\mu)}{\partial t}&\propto& \lambda_{SH}(\mu)\ .\label{eq:mixing_rge}
\eea
Thus, the light scalar mass stays small under RGE running, despite corrections from the heavy scalar $H$, due to the fixed point found in the technically natural limit $\lambda_{SH}\rightarrow0$ and $m_S^2\rightarrow0$ in Eq.~\eqref{eq:light_heavy_scalar_rge}. This reflects the enhanced Poincar\'e symmetry found in the decoupling limit. 

We note a third way to obtain a naturally light scalar: if the beta-function for $m_S$ depends on a set of couplings $\lambda_i$ such that the running of $\lambda_i$ approaches a fixed point that sends $\beta_{m_S}$ to zero, then $m_S$ will not run large. We do not know of a model that can achieve this but it appears to be a logical possibility.




\begin{thebibliography}{99}
\bibitem{Gildener:1976ai} 
  E.~Gildener,
  Phys.\ Rev.\ D {\bf 14}, 1667 (1976).

\bibitem{Susskind:1978ms} 
  L.~Susskind,
  Phys.\ Rev.\ D {\bf 20}, 2619 (1979);
  S.~Weinberg,
  Phys.\ Lett.\ B {\bf 82} 387 (1979);
  E.~Gildener,
  Phys.\ Lett.\ B {\bf 92} 111 (1980).

\bibitem{Bardeen:1995kv}
  W.~A.~Bardeen,
  FERMILAB-CONF-95-391-T.

\bibitem{Farina:2013mla} 
  M.~Farina, D.~Pappadopulo and A.~Strumia,
  JHEP {\bf 1308}, 022 (2013)
  [arXiv:1303.7244 [hep-ph]].

\bibitem{Strumia}
This point has been made by A. Strumia in recent talks, for example at the WIN 2013 workshop in Natal, Brazil.

\bibitem{Volkas:1988cm} 
  R.~R.~Volkas, A.~J.~Davies and G.~C.~Joshi,
  Phys.\ Lett.\ B {\bf 215}, 133 (1988).

\bibitem{Peccei:1977hh} 
  R.~D.~Peccei and H.~R.~Quinn,
  Phys.\ Rev.\ Lett.\  {\bf 38}, 1440 (1977);
  Phys.\ Rev.\ D {\bf 16}, 1791 (1977).

\bibitem{Weinberg:1977ma} 
  S.~Weinberg,
  Phys.\ Rev.\ Lett.\  {\bf 40}, 223 (1978);
  F.~Wilczek,
  Phys.\ Rev.\ Lett.\  {\bf 40}, 279 (1978).

\bibitem{Kim:1979if} 
  J.~E.~Kim,
  Phys.\ Rev.\ Lett.\  {\bf 43}, 103 (1979);
A.~R.~Zhitnitskii,
 Yad.\ Fiz.\ {\bf 31}, 497 (1980);
  M.~Dine, W.~Fischler and M.~Srednicki,
  Phys.\ Lett.\ B {\bf 104}, 199 (1981).

\bibitem{Preskill:1982cy} 
  J.~Preskill, M.~B.~Wise and F.~Wilczek,
  Phys.\ Lett.\ B {\bf 120}, 127 (1983).

\bibitem{Vissani:1997ys} 
  F.~Vissani,
  Phys.\ Rev.\ D {\bf 57}, 7027 (1998)
  [hep-ph/9709409].




\bibitem{Meissner:2006zh}
  K.~A.~Meissner and H.~Nicolai,
  Phys.\ Lett.\ B {\bf 648} 312 (2007) 
  [hep-th/0612165].

\bibitem{Foot:2007as}
  R.~Foot, A.~Kobakhidze and R.~R.~Volkas,
  Phys.\ Lett.\ B {\bf 655} 156 (2007)
  [arXiv:0704.1165 [hep-ph]];
  R.~Foot, A.~Kobakhidze, K.~L.~McDonald and R.~R.~Volkas,
  Phys.\ Rev.\ D {\bf 76} 075014 (2007)
  [arXiv:0706.1829 [hep-ph]].

\bibitem{Foot:2007iy}
  R.~Foot, A.~Kobakhidze, K.~L.~McDonald and R.~R.~Volkas,
  Phys.\ Rev.\ D {\bf 77} 035006 (2008)
  [arXiv:0709.2750 [hep-ph]].


\bibitem{Iso:2009ss} 
  S.~Iso, N.~Okada and Y.~Orikasa,
  Phys.\ Lett.\ B {\bf 676}, 81 (2009)
  [arXiv:0902.4050 [hep-ph]];
  S.~Iso, N.~Okada and Y.~Orikasa,
  Phys.\ Rev.\ D {\bf 80}, 115007 (2009)
  [arXiv:0909.0128 [hep-ph]];
  M.~Holthausen, M.~Lindner and M.~A.~Schmidt,
  Phys.\ Rev.\ D {\bf 82}, 055002 (2010)
  [arXiv:0911.0710 [hep-ph]].


\bibitem{Foot:2010av}
  R.~Foot, A.~Kobakhidze and R.~R.~Volkas,
  Phys.\ Rev.\ D {\bf 82} 035005 (2010)
  [arXiv:1006.0131 [hep-ph]].

\bibitem{AlexanderNunneley:2010nw} 
  L.~Alexander-Nunneley and A.~Pilaftsis,
  JHEP {\bf 1009}, 021 (2010)
  [arXiv:1006.5916 [hep-ph]];
  R.~Foot, A.~Kobakhidze and R.~R.~Volkas,
  Phys.\ Rev.\ D {\bf 84}, 075010 (2011)
  [arXiv:1012.4848 [hep-ph]];
  R.~Foot and A.~Kobakhidze,
  arXiv:1112.0607 [hep-ph];
  K.~Ishiwata,
  Phys.\ Lett.\ B {\bf 710}, 134 (2012)
  [arXiv:1112.2696 [hep-ph]];
  J.~S.~Lee and A.~Pilaftsis,
  Phys.\ Rev.\ D {\bf 86}, 035004 (2012)
  [arXiv:1201.4891 [hep-ph]];
  N.~Okada and Y.~Orikasa,
  Phys.\ Rev.\ D {\bf 85}, 115006 (2012)
  [arXiv:1202.1405 [hep-ph]];
  S.~Iso and Y.~Orikasa,
  PTEP {\bf 2013}, 023B08 (2013)
  [arXiv:1210.2848 [hep-ph]];
  C.~Englert, J.~Jaeckel, V.~V.~Khoze and M.~Spannowsky,
  JHEP {\bf 1304}, 060 (2013)
  [arXiv:1301.4224 [hep-ph]];
  M.~Heikinheimo, A.~Racioppi, M.~Raidal, C.~Spethmann and K.~Tuominen,
  arXiv:1304.7006 [hep-ph];
  M.~Heikinheimo, A.~Racioppi, M.~Raidal, C.~Spethmann and K.~Tuominen,
  Nucl.\  Phys.\ B {\bf 876} 201 (2013)
  [arXiv:1305.4182 [hep-ph]];
  T.~Hambye and A.~Strumia,
  arXiv:1306.2329 [hep-ph];
  I.~Bars, P.~Steinhardt and N.~Turok,
  arXiv:1307.1848 [hep-th];
  M.~Heikinheimo, A.~Racioppi, M.~Raidal and C.~Spethmann,
  arXiv:1307.7146 [hep-ph];
  C.~D.~Carone and R.~Ramos,
  arXiv:1307.8428 [hep-ph];
  A.~Farzinnia, H.~-J.~He and J.~Ren,
  arXiv:1308.0295 [hep-ph];
  Y.~Kawamura,
  arXiv:1308.5069 [hep-ph];
  V.~V.~Khoze,
  arXiv:1308.6338 [hep-ph];
  E.~Gabrielli, M.~Heikinheimo, K.~Kannike, A.~Racioppi, M.~Raidal and C.~Spethmann,
  arXiv:1309.6632 [hep-ph].


\bibitem{Shaposhnikov:2008xi} 
  M.~Shaposhnikov and D.~Zenhausern,
  Phys.\ Lett.\ B {\bf 671}, 162 (2009)
  [arXiv:0809.3406 [hep-th]];
  D.~Blas, M.~Shaposhnikov and D.~Zenhausern,
  Phys.\ Rev.\ D {\bf 84} (2011) 044001
  [arXiv:1104.1392 [hep-th]];
  R.~Armillis, A.~Monin and M.~Shaposhnikov,
  arXiv:1302.5619 [hep-th];
  F.~Gretsch and A.~Monin,
  arXiv:1308.3863 [hep-th].


\bibitem{Tamarit:2013vda} 
  C.~Tamarit,
  arXiv:1309.0913 [hep-th].

\bibitem{Coleman:1973jx}
  S.~R.~Coleman and E.~J.~Weinberg,
  Phys.\ Rev.\ D {\bf 7} 1888 (1973).

\bibitem{Gildener:1976ih}
  E.~Gildener and S.~Weinberg,
  Phys.\ Rev.\ D {\bf 13} 3333 (1976).

\bibitem{'tHooft:1979bh}
  G.~'t Hooft,
  NATO Adv.\ Study Inst.\ Ser.\ B Phys.\  {\bf 59} 135 (1980).




\bibitem{Antoniadis:1984kd} 
  I.~Antoniadis and N.~C.~Tsamis,
  Phys.\ Lett.\ B {\bf 144}, 55 (1984);
  C.~Wetterich,
  Nucl.\ Phys.\ B {\bf 302}, 668 (1988).







\end{thebibliography}
\end{document}